\documentclass{aa}  
\usepackage{color}
\usepackage{natbib}
\usepackage{graphicx}
\usepackage{amsmath}
\usepackage{txfonts}
\usepackage{physics}
\newcommand{\vect}[1]{\vb{#1}}
\begin{document} 

\title{Towards inertial-mode helioseismology:\\ Direct sensing of solar rotation at  $75^\circ$ latitude and   $0.8 R_\odot$}
   \author{Prithwitosh Dey\inst{1}
   \and{Laurent Gizon\inst{1,2,3}\thanks{Corresponding author. email: gizon@mps.mpg.de}}
   \and{Yuto Bekki\inst{1}}}
   \institute{Max-Planck-Institut f{\"u}r Sonnensystemforschung, Justus-von-Liebig-Weg 3, 37077 G{\"o}ttingen, Germany
    \and
        Institut f{\"u}r Astrophysik und Geophysik, Georg-August-Universtät Göttingen, Friedrich-Hund-Platz 1, 37077 G{\"o}ttingen, Germany
        \and
        Center for Astrophysics and Space Science, NYUAD Institute,
New York University Abu Dhabi, Abu Dhabi, United Arab Emirates
}

\date{Received XXX; accepted YYY}

\abstract
   {Solar internal rotation at high latitudes is poorly constrained by acoustic-mode helioseismology. Global  inertial modes observed on the Sun  are highly sensitive to solar differential rotation  and may provide new diagnostics of rotation in these regions.}
   {We aim to constrain  solar rotation with the measured frequency of the $m=1$ high-latitude inertial mode, starting from the {HMI/SDO} reference rotation profile given by p-mode helioseismology for 2010--2024.}
  {
  Using a validated and accurate eigenvalue solver, we compute the perturbation to the  mode frequency resulting from localised changes in the differential rotation rate throughout the solar interior. 
}
   {
We find that the linear sensitivity kernel of the $m=1$ high-latitude mode peaks at latitude $75^\circ$ and radius  $0.8 R_\odot$, with full widths of $7^\circ$ and $0.13 R_\odot$. 
  From the observed mode frequency in the Carrington frame,   $-87.9 \pm 1.9$ nHz (retrograde, averaged over 2010--2024),  
   we infer that the solar rotation rate  near this location is $365.3\pm2.0$~nHz, which exceeds  the reference p-mode estimate by $8.1$~nHz.
   Additionally, we propose a latitudinally smooth, radially independent modification to the rotation rate at high latitudes beyond the linear (small-perturbation) regime.
   }
   {This work demonstrates that  individual  inertial modes can provide direct constraints on rotation in the bulk of the solar convection zone, well below the surface, representing the first example of spatially resolved inertial-mode helioseismology.
   }

   \maketitle

\section{Introduction}

Internal differential rotation is crucial for understanding the Sun's dynamo and angular momentum transport mechanisms \citep[e.g.,][]{kapyla2023}. Our current knowledge of the Sun's  differential rotation comes primarily from p-mode helioseismology, which exploits the frequency splittings of  modes  with different azimuthal orders
\citep[see review  article by][]{thompson2003}. The Sun's convection zone (CZ), which constitutes the outer 30\% of the Sun, rotates differentially, varying with both radius and latitude, while the radiative interior rotates almost rigidly. The tachocline exists at the transition between these two zones,  \citep[e.g.,][]{Spiegel1992}, as well as  a near-surface shear layer in the outermost 5\% of the solar radius \citep[e.g.,][]{thompson1996}.

Despite these remarkable achievements, p-mode helioseismology still faces important limitations. It provides limited information about the Sun’s rotation below  $\sim0.2R_\odot$ \citep[e.g.][]{gizon1997} and in regions within $\sim 20^\circ$ of the rotation axis \citep[e.g.][]{schou1998}. This limitation arises because very few p-modes have appreciable kinetic energy density in these regions, as their sensitivity kernels peak near the solar surface. As a consequence, inversions of frequency splittings obtained using different methods can differ significantly in the polar regions \citep[][their figure~7]{thompson2003}.

Solar inertial modes are low-frequency global modes of oscillation that are restored by the Coriolis force. These are retrograde modes with a strong toroidal component and periods on time scales comparable to the solar rotation period ($\sim 25$~days at the equator). {Maps of horizontal flows near the solar surface have been measured for more than a decade using p-mode local helioseismology and local correlation tracking applied to data from the Helioseismic and Magnetic Imager aboard the Solar Dynamics Observatory (HMI/SDO) and Global Oscillations Network Group (GONG).} These long-duration observations allow for the unambiguous detection and characterization of solar inertial modes \citep{loeptien2018, gizon2021, hanson2022}.

\begin{figure}
    \centering
\includegraphics[width=0.8\linewidth]{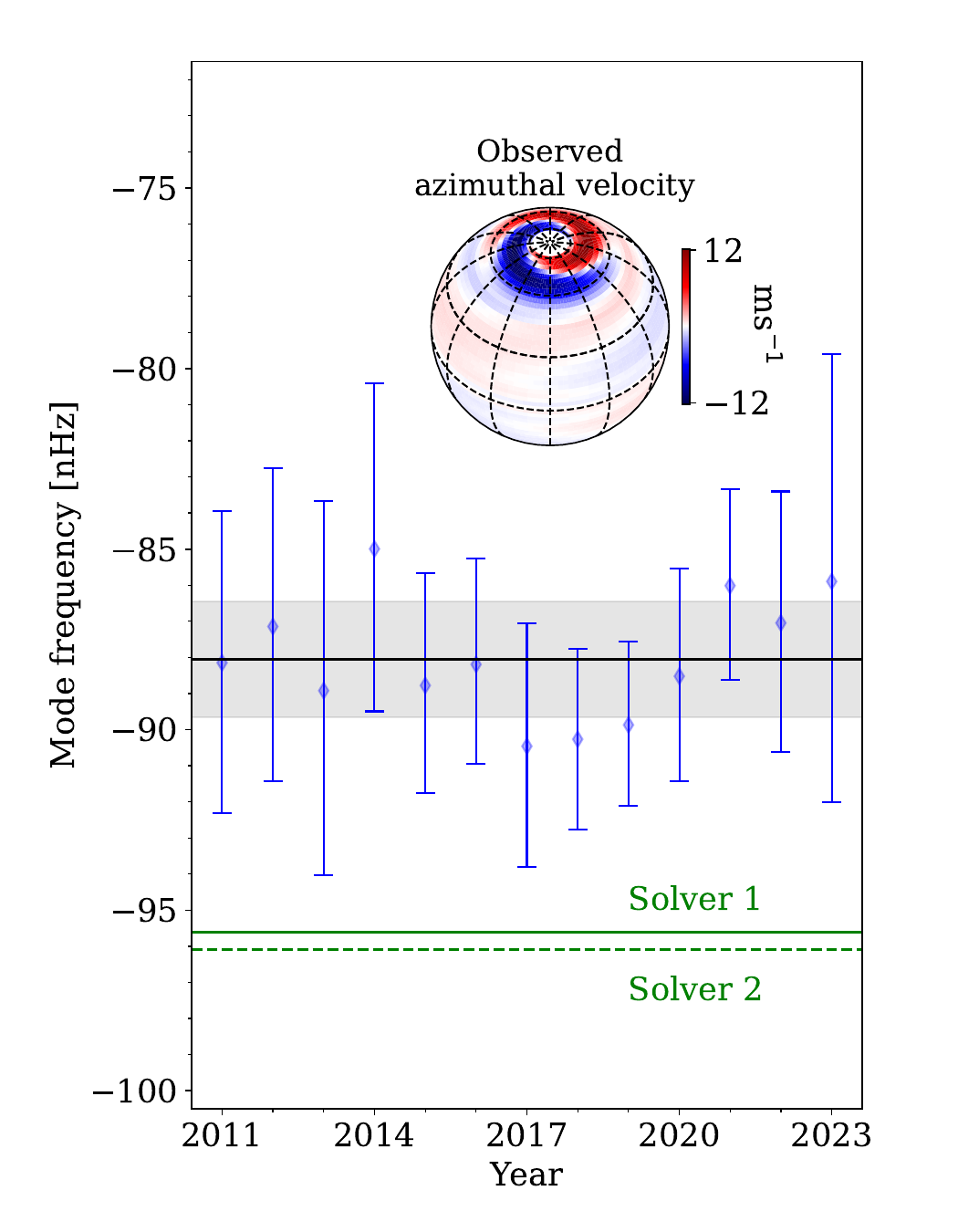}
    \caption{Observed HL1 frequencies in the Carrington reference frame, extracted from HMI Dopplergrams at a cadence of one year with a three-year sliding window \citep[blue points with error bars,][]{liang2024}. The mean frequency $\omega_{\rm obs}/2\pi = -87.9$~nHz is indicated by the black horizontal line, and the gray shaded region shows the $\pm 1\sigma$ range ($\pm 1.9$~nHz). 
    The frequency uncertainty was estimated from measurements over independent time intervals centered on 2012, 2015, 2018, and 2021, thereby accounting for correlations introduced by the sliding window. The green solid line marks the real part of the eigenfrequency computed using eigenvalue Solver~1 ($-95.6$~nHz) and the green dashed line marks the same for Solver~2 ($-96.1$~nHz). The azimuthal velocity of the observed mode at the surface is shown as an inset \citep[adapted from][]{gizon2021}.}
\label{fig:time_variation}
\end{figure}

In this Letter we focus on the {high-latitude mode with azimuthal wavenumber $m=1$ and north-south symmetric radial vorticity (hereafter the `HL1' mode)}. This  mode is observed to have  the largest amplitude on the surface of the Sun, in the range 5--25 m/s at latitudes above $60^\circ$ \citep{gizon2021}. It has been consistently detected in multiple data sets and is also present in the Mount Wilson Observatory  Dopplergrams spanning the last five sunspot cycles \citep{ulrich2001,liang2024}. The frequency of the HL1 mode is known to a high degree of precision. Over the period of HMI observations, its mean frequency in the Carrington frame is $\omega_{\rm obs}/2\pi = -87.9\pm 1.9$~nHz (Fig.~\ref{fig:time_variation}).

Because their physics is intrinsically linked to rotation, inertial modes are particularly well suited for diagnosing differential rotation \citep[see][for the case of Rossby modes]{goddard2020}. Linear eigenmode computations indicate that the kinetic energy density of the HL1 mode peaks inside the convection zone \citep{gizon2021,bekki2022a}, making it a very promising candidate for directly probing rotation in the solar interior.

\label{reference_model}

\section{Mismatch between model and observed  frequencies of $HL1$  inertial mode}

In order to place constraints on the solar internal rotation using the observed frequency of the HL1 mode, we  need to solve the forward problem, that is, to compute the mode frequency for a given prescription of the solar differential rotation (see Appendix~A).
Let us denote the solar angular velocity measured in an inertial frame by $\Omega(r,\theta)$, where $r$ is the radius and $\theta$ the colatitude ($\theta=0$ at the north pole). We adopt the reference differential rotation rate $\Omega_{\rm ref}(r,\theta)$ derived from p-mode helioseismology \citep{larson2018}. In the Carrington frame rotating at angular frequency $\Omega_{\rm Carr}/2\pi = 456.0$~nHz, we linearise the equations of momentum, entropy, and continuity about a standard solar model \citep[Model S,][]{christensen1996}.
The computational domain extends radially over the range 
$0.65R_\odot < r < 0.985R_\odot$.  We assume   perturbations proportional to 
$\exp({\rm i}m\phi-{\rm i}\omega t)$, where $m=1$ and  $\phi$ is the longitude.
 Under horizontal stress-free boundary conditions, the system  yields a discrete spectrum of complex eigenvalues $\omega$ and corresponding eigenvectors. 
We use an eigenvalue solver based on \citet{bekki2022a} to compute the eigenmodes in the inertial frequency range 
(Solver~1; see Appendix~\ref{sec:Computation}). 
This solver was validated through a code-to-code benchmark comparison with the independent implementation of \citet{mukhopadhyay2025} (Solver~2; see Appendix~\ref{sec:Computation}). 
The $m=1$ high-latitude eigenmode is identified in the spectrum as the only self-excited low-frequency mode ($\Im{\omega} > 0$) that exhibits north-south symmetric radial vorticity.
Using Solver 1, we obtain a model eigenfrequency of $\omega_{\rm ref}/2\pi=-95.6+3.9\, \textrm{i}$~nHz in the Carrington  frame. The real part of $\omega_{\rm ref}$ thus differs from the observed  $\omega_{\rm obs}$ by four standard deviations (see Fig.~\ref{fig:time_variation}).  The velocity  eigenfunction at the surface, however, is  similar to the observation, with power confined to a similar latitudinal range above $60^\circ$  and a qualitatively similar spiral pattern. 

{We performed a set of eigenvalue computations that indicate that this discrepancy  is likely due  to an incorrect reference differential rotation profile. This discrepancy cannot be fixed by varying the other parameters of the background solar model, such as the viscous or thermal diffusivities or the superadiabaticity. 
In particular, we computed eigenmodes around the commonly used value of turbulent diffusivity in the convection zone $\nu_{\rm CZ}=10^{12}\ \rm cm^2s^{-1}$. Within the range $10^{11}\ \rm cm^2s^{-1} \leq \nu_{\rm CZ} \leq 2\times10^{12}\ \rm cm^2s^{-1}$, we find that the eigenfrequency  remains within the bounds $5.0~{\rm nHz}<(\omega_{\rm obs}- \Re{\omega})/2\pi<10.0~{\rm nHz}$,
indicating that the diffusion cannot be used as a tunable parameter to bring the eigenfrequency within observational errorbars. Further, we find that the superadiabaticity {$(\delta = \nabla-\nabla_{\rm ad})$} affects the eigenfrequency only weakly, by at most $3$~nHz when the superadiabatic gradient is varied in the range $|\delta| < 5\times 10^{-7}$.}
This motivates focusing on the background rotation rate as the primary physical parameter of the model that can be adjusted to bring the model into agreement with the observations, while keeping fixed   $\delta=0$ and $\nu_{\rm CZ} = 10^{12}$~cm$^2$s$^{-1}$.

\begin{figure}
    \centering
\includegraphics[width=0.8\linewidth]{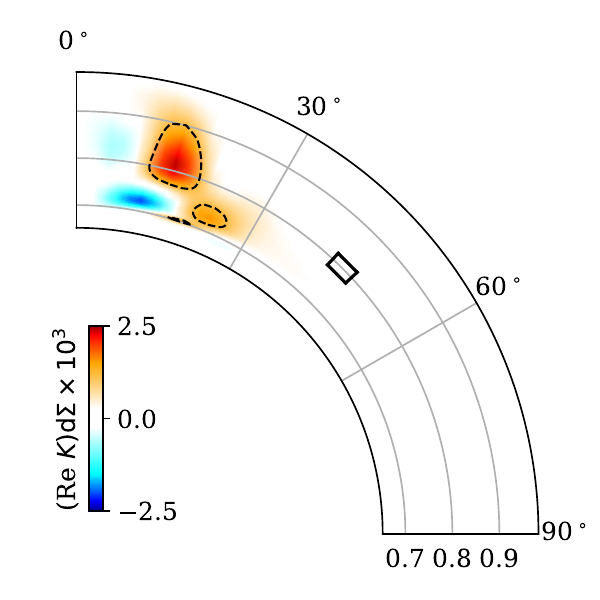}
    \caption{Spatial sensitivity of the frequency of the HL1 mode  to localised changes in the background rotation rate. The kernel $K(r,\theta)$ is symmetric across the equator. Gridsizes  ${\rm d}r =2.3$~Mm and ${\rm d}\theta= 2^\circ$ have been used for this calculation. For reference, a patch of area $10 {\rm d} r \times 2r{\rm d}\theta= 20 {\rm d} \Sigma$  is outlined with a solid black contour. The dashed black contour corresponds to the value of the kernel at half maximum.}
    \label{fig:Kernel}
\end{figure}

\section{Sensitivity of $HL1$ mode frequency to the local rotation rate}

The linear sensitivity of the mode frequency to a spatial change in rotation is specified through the integral equation
\begin{equation}
    \delta\omega = \int_0^\pi\int_0^{R_\odot}K(r, \theta)\ \delta\Omega(r, \theta)\ r{\rm d}r{\rm d}\theta,
\end{equation}
where  $\delta\omega=\omega-\omega_{\rm ref}$ is the perturbation to the (complex) eigenfrequency that results from a  small perturbation in the rotation rate,  $\delta\Omega=\Omega-\Omega_{\rm ref}$, and $K$ is the (complex) sensitivity kernel. The discretized version of this equation is
\begin{equation}
    \delta\omega = \sum_{i,j} K(r_i, \theta_j)\ \delta\Omega(r_i, \theta_j)\ {\rm d}\Sigma,
\end{equation}
where ${\rm d}\Sigma = r\,{\rm d}r\,{\rm d}\theta$ is the  element of surface in the meridional plane, expressed in terms of the radial and latitudinal grid spacings, ${\rm d}r = 2.3$  Mm and ${\rm d}\theta = 2.0^\circ$, used in our calculations.
The most straightforward method for computing the sensitivity kernel is numerical. For a given spatial pixel $(r_i, \theta_j)$ where the kernel is to be evaluated, we consider a rotation profile that is perturbed only at that pixel,
\begin{equation}
\delta\Omega(r_p, \theta_q) = \epsilon\ \delta_{pi} \delta_{qj},
\end{equation}
where the grid $(r_p, \theta_q)$ spans the full numerical domain. Here, $\delta_{pi}$ and $\delta_{qj}$ are Kronecker delta functions. The perturbation amplitude $\epsilon$ must be sufficiently small to ensure that the resulting change in oscillation frequency remains in the linear regime. We choose $\epsilon/2\pi=1.0$~nHz.
The latitudinal entropy gradient is also perturbed numerically to maintain thermal wind balance.
Using Solver~1, we compute the eigenfrequency of the high-latitude mode, $\omega_\epsilon(r_i,\theta_j)$, for this  perturbed system. 
The sensitivity kernel at pixel $(i,j)$ is then given by
\begin{equation}
    K(r_i, \theta_j)\ {\rm d}\Sigma 
    =[{\omega_\epsilon(r_i,\theta_j)-\omega_{\rm ref}}]/{\epsilon}.
\end{equation}
{The real part of the kernel, presented in Fig.~\ref{fig:Kernel}, shows a prominent peak centered at $(r_0,\theta_0)=(0.82R_\odot,15^\circ)$, and two smaller sidelobes, one positive and one negative, at $0.72R_\odot$.}
The main peak of the kernel has full widths at half maximum of $7^\circ$ in latitude and $0.13 R_\odot$ in radius, as shown in the cuts in Figs.~\ref{fig:Kernel_radial_cut} and~\ref{fig:Kernel_latitudinal_cut}. We find almost no sensitivity in the radiative interior.

Because the kernel is reasonably well localized in both radius and latitude, the inferred frequency shift can be interpreted as providing a direct estimate of the local rotation perturbation at location $(r_0, \theta _0)$. 
Specifically, the observed frequency difference can be related to the local change in rotation rate through the integral of the kernel over the meridional plane. The total integral of the kernel over the meridional plane, $\sum_{ij} K(r_i,\theta_j) {\rm d}\Sigma=0.95+0.03\,\mathrm{i}$,  has a real part which is close to unity which simplifies the interpretation of the data.\footnote{The total integral of the kernel can also be computed directly by perturbing the reference rotation rate by a constant value $\epsilon$; this yields $\delta\omega/\epsilon = 0.96 + 0.03\,\mathrm{i}$, confirming the accuracy of the eigenvalue solver}.
We do not attempt a quantitative interpretation of the imaginary part, since there is presently no established method to relate the imaginary part of the mode frequency to an observable   \citep[such as the mode amplitude, see][for a preliminary weakly nonlinear approach in 2D]{mushtaq2026}.
Using the constraint $(\omega_{\rm obs}-\Re{\omega_{\rm ref}})/2\pi=7.7 \pm 1.9$~nHz, we estimate the perturbation to the rotation rate  near $(r_0, \theta _0)$ as 
\begin{align}
\label{eq.simpleinversion}
    \delta \Omega(r_0, \theta _0)/2\pi \approx  (7.7 \pm 1.9)/0.95\  \text{\rm nHz} = 8.1 \pm 2.0\ \text{\rm nHz} . 
\end{align}
This small but significant enhancement of the rotation rate  relative to the p-mode reference value ($\Omega_{\rm ref}(r_0, \theta _0)/2\pi = 357.2$~nHz), shows that the HL1 mode provides a sensitive probe of differential rotation at high latitudes in the deep convection zone.

The above constraint is compared in Fig.~\ref{fig:rotation_updated} with a p-mode rotation estimate at $(r_0,\theta_0)$ obtained using optimally localized averaging \citep[OLA;][]{howe2009}. This method combines p-mode frequency splittings to produce a resolution kernel localized at the target position \citep{Pijpers1992,pijpers1994}.  A significant discrepancy is observed, at a colatitude where OLA inversions are most uncertain.  Remarkably, a single inertial mode  yields a kernel with  comparable latitudinal extent.

\begin{figure}
    \centering      \includegraphics[width=0.8\linewidth]{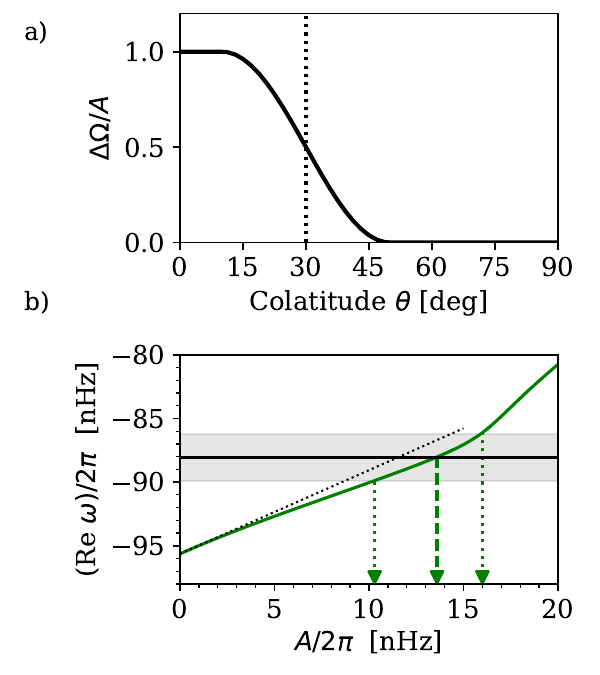}
    \caption{Smooth high-latitude modification to the background rotation rate, $\Delta\Omega(\theta; A)$, and its effect on the HL1 mode frequency.
    (a) Functional form of  $\Delta\Omega(\theta;A)$ as defined by Eq.~(\ref{eq:modification}), symmetric across the equator. (b) Variation of the real part of the mode frequency with $A$, using Solver~1 (green curve). By construction $\omega=\omega_{\rm ref}$ at $A=0$. Agreement with the observed eigenfrequency (gray shaded region) is obtained for $10.3~{\rm nHz} < A/2\pi < 16.0~{\rm nHz}$ (range between the green dotted arrows). The value of $A/2\pi=13.6$~nHz corresponding to the best fit is marked by the dashed green arrow. The dotted black line shows the linear approximation  Eq.~(\ref{eq:linear_variation}). }\label{fig:profiles_frequency}
\end{figure}

\section{Parametric model for rotation correction}
\label{sec:parametric_model}
To study the effect of a smooth modification to the rotation rate at high latitudes, we define a simple parametric correction that depends only on colatitude:
\begin{equation}
\label{eq:modification}
\Delta\Omega(\theta; A) = \begin{cases}
A & \theta < 10^\circ, \\
A  [1 + \sin\left( 4.5 (30^\circ-\theta)    \right) ]/2 &  
  10^\circ\leq\theta < 50^\circ, \\
0 & 50^\circ \leq \theta < 90^\circ.
\end{cases}
\end{equation}
This rotation perturbation  is    symmetric across the equator. In the above equation, the profile reaches a maximum amplitude $A$ at the pole and has a width of $30^\circ$ (see Fig.~\ref{fig:profiles_frequency}a).
This perturbation is radially constant, and it therefore also affects the rotation rate in the radiative zone. However, we showed in the previous section that the mode is insensitive to rotation in the radiative zone and we choose not to add a radial dependence of the profile for simplicity.

For infinitesimal perturbations in the angular velocity parametrised as above (i.e., as ${A\rightarrow 0}$), the response of the mode frequency  is
\begin{equation}
\label{eq:linear_variation}
 \delta \omega(A) \approx \int_\Sigma K(r,\theta)\ \Delta \Omega(\theta; A)\ {\rm d}\Sigma = (0.65 +  0.20\,\mathrm{i})\ A  .
\end{equation}
In the linear regime, 
 $A \approx (\omega_{\rm obs}-\Re{ \omega_{\rm ref} })/0.65$. However, the range of validity of this linear approximation is not yet established, motivating the direct numerical computations of $\Re{\delta \omega(A)}$ for finite-amplitude values of $A$.

\begin{figure}
    \centering    
    \includegraphics[width=\linewidth]{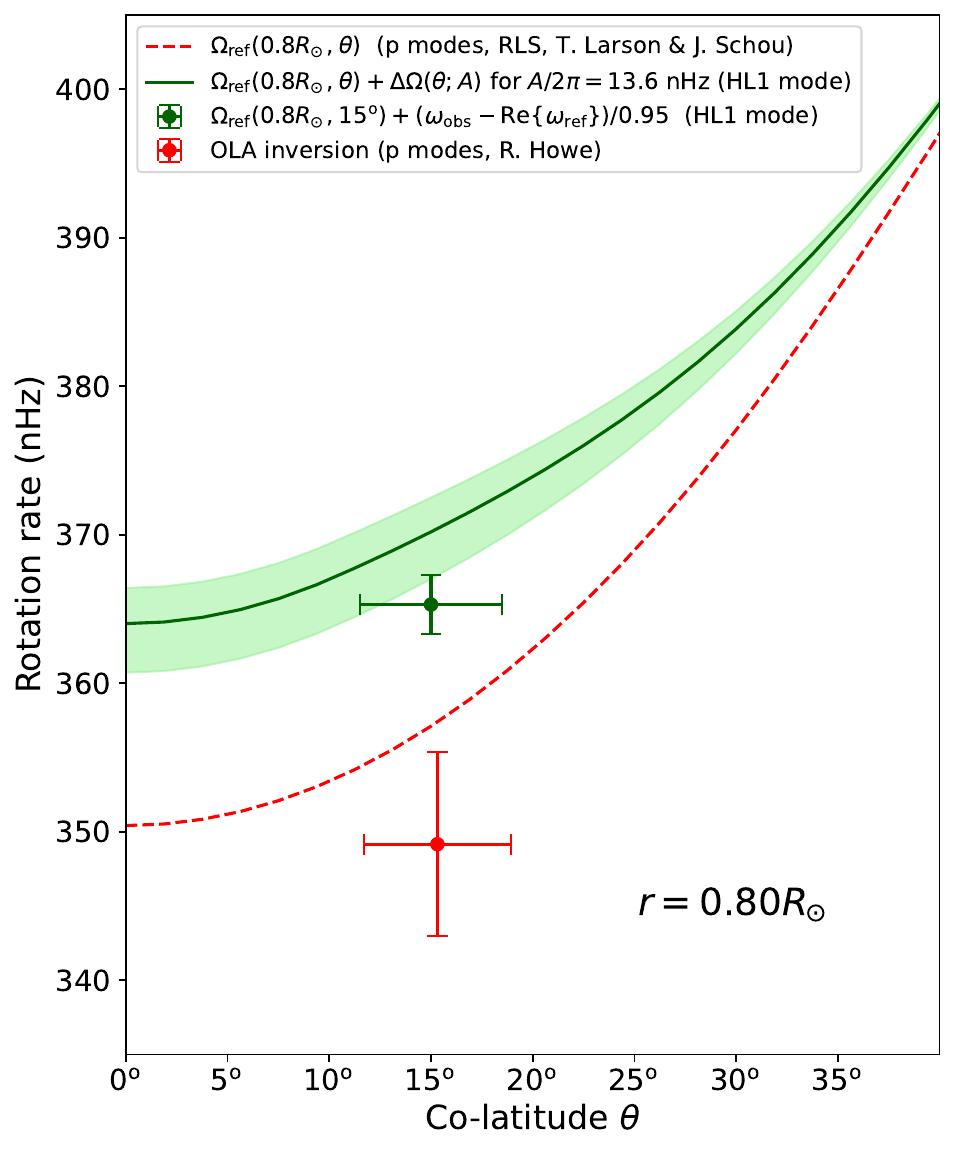}
    \caption{Constraints on high-latitude solar rotation rate at $r = 0.8\,R_\odot$. The green cross shows the constraint from the HL1 mode frequency (2010--2024), according to Eq.~(\ref{eq.simpleinversion}), with the horizontal bar indicating the width of the kernel.
    The green curve shows the rotation rate $\Omega_{\rm ref}(r,\theta) + \Delta\Omega(\theta; A)$  for  $A/2\pi = 13.6$~nHz  and the green shaded region indicates the associated uncertainty (see Fig.~\ref{fig:profiles_frequency}). 
    The p-mode reference rotation rate $\Omega_{\rm ref}(0.8 R_\odot, \theta)$  is given by the red dashed curve. 
    The red cross shows a p-mode OLA estimate at  target location $(0.8\,R_\odot, \theta = 15^\circ)$ obtained from  yearly HMI inversions \citep{howe2009} averaged over 2010--2023. The horizontal bar represents the width of the averaging kernel, while the vertical bar represents the random error.
     }
\label{fig:rotation_updated}
\end{figure}

 We recompute the eigenfrequency of the mode  with the modified rotation rate $\Omega(r, \theta) = \Omega_{\rm ref}(r, \theta) + \Delta\Omega(\theta; A)$ for  values of $A/2\pi$ ranging from $0$ to $20$~nHz. In all cases, the  latitudinal entropy gradient is recomputed such that the thermal wind balance is maintained. The variation of the real part of the eigenfrequency  with $A$ is presented in Fig.~\ref{fig:profiles_frequency}b. 
The relationship between the real part of the eigenfrequency and $A$ is roughly linear for $A/2\pi \lesssim 16$~nHz.
For $A/2\pi$ between $10.3$~nHz and $16.0$~nHz, the real part of the eigenfrequency is within $\pm 1\sigma$ of the observed frequency and the imaginary part of the frequency is positive (the mode is linearly unstable).
A profile of the modified rotation rate at $r = 0.8\,R_\odot$ is shown in Fig.~\ref{fig:rotation_updated} for $A/2\pi = 13.6^{+2.4}_{-3.3}$~nHz, giving a  polar rotation rate of $364.0$~nHz.

\section{Conclusion}

We found that the rotation kernel of the HL1 mode frequency is highly localized in the convection zone near $r=0.8R_\odot$ and  $\theta=15^\circ$. Without performing any inversion, the integrated HL1 kernel implies a local rotation rate of approximately $365.3\pm2.0$~nHz at this location, that is, $8.1\pm2.0$~nHz above the reference  estimate from p-mode helioseismology \citep[RLS,][]{larson2018}, and also above  a p-mode OLA inversion \citep{howe2009}. Using the smooth parametric profile prescribed in Section~\ref{sec:parametric_model}, the inferred enhancement reaches $13.6\pm3$~nHz at the pole relative to the reference model.
{These estimates are absolute measurements in the sense that we used the frequency of a single retrograde inertial mode. In contrast, acoustic modes infer the rotation rate using a differential measurement between prograde and retrograde modes, which reduces dependencies on the background model.}

The frequencies of many other solar inertial modes have been measured \citep{gizon2021}, as well their variations with the solar cycle \citep{waidele2023,lekshmi2026}. 
 Future work will aim to constrain the solar differential rotation at multiple positions in the convection zone using all modes simultaneously by solving an inverse problem. In this respect, the current situation is reminiscent of the early days of helioseismology, when \citet{JCD1976} foresaw the potential of p modes as probes of the solar interior.
 The validated eigenvalue solvers described in this paper will be valuable tools in this respect, in particular for interpreting the full spectrum of inertial modes and for assessing the contribution of time-varying zonal flows to their temporal variations
\citep[see][]{goddard2020}. 
 
\begin{acknowledgements}
{PD and LG designed and performed research, PD and YB updated an earlier version of Solver~1, PD wrote the first draft of the paper and all authors contributed to the final manuscript. 
The authors acknowledge the essential contribution of Suprabha Mukhopadhyay, which enabled the cross-validation of Solver~1 with Solver~2. We thank Zhi-Chao Liang for making available the  HL1 mode frequencies shown in Fig.~\ref{fig:time_variation} and Rachel Howe  for providing the p-mode OLA rotation inversion shown in Fig.~\ref{fig:rotation_updated}. We acknowledge useful discussions with Damien Fournier and Robert Cameron.
We acknowledge partial financial support from ERC Synergy Grant WHOLESUN (grant agreement No. 810218) to LG. PD is a member of the International Max Planck Research School for Solar System Science at the University of Göttingen. }
\end{acknowledgements}

\bibliographystyle{aa} 
\bibliography{ref}

\appendix

\section{Computation of eigenmodes of a rotating solar model}
\label{sec:Computation}
\subsection{Linearized equations of motion}

We work in the Carrington frame of reference, rotating at frequency $\Omega_{\rm Carr}/2\pi=456.0$~nHz with respect to an inertial frame.
We use the linearised forms of the momentum, entropy, and continuity equations, where perturbations in velocity ($\vect{v}'$), pressure ($p'$), density ($\rho'$), and entropy ($s'$) are proportional to $\exp[{\rm i}(m\phi - \omega t)]$, where $\phi$ is the longitude and $\omega$ is the angular frequency of the mode of oscillation, to obtain
\begin{align}
 {\rm i}\omega \vect{v}'   =\ &  
 \frac{\nabla p'}{\rho}
 +\frac{\rho'}{\rho}g \hat{\vect{r}} 
 -2{\rm{\bold v}}'\times(\Omega\hat{\vect{z}})
 + r\sin\theta (\vect{v}'\cdot\nabla\Omega) \hat{\boldsymbol{\phi}}
 \nonumber\\
& -\frac{\nabla\cdot\boldsymbol{\mathcal{D}}}{\rho} 
+{\rm i}m(\Omega-\Omega_{\rm Carr})\vect{v}',
\label{eq:motion} \\
{\rm i}\omega s'  =\ &  \vect{v}'\cdot\nabla s -\frac{\nabla\cdot(\kappa \rho T \nabla s')}{\rho T} + {\rm i}m(\Omega-\Omega_{\rm Carr})s',\label{eq: entropy} \\
{\rm i}\omega \rho' =\ & \nabla\cdot(\rho\vect{ v}') + {\rm i}m(\Omega-\Omega_{\rm Carr})\rho'. \label{eq:mass}
\end{align}
In the above equations,  the unprimed variables ($\rho$, $p$, $g$, $s$, $T$) refer to the background reference state, and $\kappa$ is the thermal diffusivity. 
The unit vectors $\hat{\vect{r}}$, $\hat{\boldsymbol{\phi}}$, and $\hat{\vect{z}}$  point in the radial, longitudinal, and rotation-axis directions, respectively.
An ideal gas equation of state is used to close the equations. The perturbations are complex functions of radius and colatitude.  The components of the viscous stress tensor $\boldsymbol{\mathcal{D}}$ are given by
\begin{equation}
\vect{\mathcal{D}}_{ij}=\rho\nu_{\rm turb}\left(S_{ij}-\frac{2}{3}\left(\delta_{ij}\nabla\cdot\vect{v}'\right)\right),
    \label{eq:viscosity}
\end{equation}
where the $S_{ij}$ are the components of the deformation tensor in spherical coordinates \citep{fan2014} and $\nu_{\rm turb}$ is the turbulent viscosity.

\subsection{Boundary conditions}
Complemented by boundary conditions, the above equations constitute an eigenvalue problem. 
We choose the computational domain to span $0.65 R_\odot$ to $0.985 R_\odot$ in radius.
We use impenetrable horizontal stress-free boundary conditions for the velocity at the top and bottom boundaries, and assume there is no entropy flux at either boundary:
\begin{equation}
    v'_r=0, \quad \frac{\partial }{\partial r}\left(\frac{v'_\theta}{r}\right)=0, \quad\frac{\partial }{\partial r}\left(\frac{v'_\phi}{r}\right)=0, \quad  \frac{\partial s'}{\partial r}=0 .
\end{equation}
Further, we assume the perturbations $v_r'$, $v_\phi'$, $\rho'$, and $s'$ to vanish at the poles ($\theta=0$ and $\pi$). For the specific case of $m=1$ (our case), we require  $\partial {v}_\theta'/\partial\theta = 0$. This  allows for a polar crossing flow.
With these boundary conditions, Eqs.~(\ref{eq:motion})--(\ref{eq:mass}) form an eigenvalue problem, which is non-singular because of the viscous stresses Eq.~(\ref{eq:viscosity}). The spectrum of eigenvalues $\omega$ is discrete.

\subsection{Reference solar model and reference differential rotation}

\begin{figure}
    \centering
\includegraphics[width=0.9\linewidth]{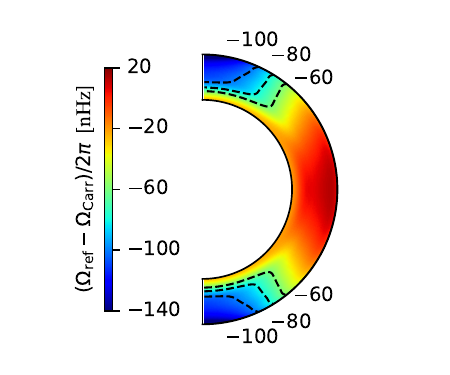}
    \caption{The reference solar angular velocity in the Carrington frame, $\Omega_{\rm ref}-\Omega_{\rm Carr}$, from p-mode helioseismology \citep{larson2018}. This meridional cut shows the computational domain for the eigenvalue solvers,  $0.65<r/R_\odot<0.985$.}
    \label{fig:rotation profile}
\end{figure}

We use the {background} density, pressure, and gravitational acceleration  from the standard solar model~S \citep{christensen1996}. 
{The reference internal rotation is provided by p-mode helioseismology \citep{larson2018}  applied to data from HMI/SDO.}
The time series of the angular velocity $\Omega(r,\theta; t)$ is derived from successive data segments of length 72~days and 36 $a$-coefficients, a standard product available for download
at the Joint Science Operations Center (JSOC) website under the filename \texttt{hmi.v\_sht\_2drls}.
We adopt as the background rotation rate the mean profile from 30 April 2010 to 16 February 2024 and denote it by $\Omega_{\rm ref}(r,\theta)$, see Fig.~\ref{fig:rotation profile}.

We need to specify $\nabla s$ in Eq.~(\ref{eq: entropy}). Invoking the thermal wind balance for the latitudinal derivative of the entropy \citep{rempel2005}, we obtain
\begin{equation}
    {\rm\bold{v}'}\cdot\nabla s = \frac{ c_p \delta}{H_p}v_r'+\frac{c_p r\sin\theta}{rg}
    \frac{\partial (\Omega_{\rm ref}^2)}{\partial z}v'_\theta,
\end{equation}
{where $c_p$ is the specific heat at constant pressure, $H_p$ is the pressure scale height, and $\delta=\nabla-\nabla_{\rm ad}$ is the superadiabatic temperature gradient.}
Because the convection zone is nearly adiabatically stratified, we set $\delta = 0$ there and smoothly transition in the radiative interior ($r/R_\odot < 0.70$) to the value of $\delta(r)$ from  Model~S.

We also need to specify $\nu_{\rm turb}$ in Eq.~(\ref{eq:viscosity}). We adopt a turbulent viscosity profile in the form of a step function:
\begin{equation}
    \nu_{\rm turb}(r) =\nu_{\rm RZ}+\frac{\nu_{\rm CZ}-\nu_{\rm RZ}}{2}\left[1+\tanh(\frac{r/R_\odot - 0.7}{0.015})\right], 
\end{equation}
where $\nu_{\rm RZ}=10^9~{\rm cm^2\,s^{-1}}$ and $\nu_{\rm CZ}=10^{12}~{\rm cm^2\,s^{-1}}$ are representative values of  the turbulent viscosities in the radiative and convection zones. We use the same profile for the thermal diffusivity.

\subsection{Two eigenvalue solvers}
\label{sec:validation}
Numerical eigenmodes may depend on the computational methods and codes.
Therefore, we considered two independent codes, which use different discretization methods {to solve the same problem}. The first eigenvalue solver, referred to as `Solver~1', is based on the finite-difference solver of \citet{bekki2022a}. The second solver, `Solver~2', is the Dedalus-based spectral solver described by \cite{mukhopadhyay2025}.
Resolution-based convergence tests were performed to ensure numerically accuracy.

\subsection{Solver~1 based on \citet{bekki2022a}}
{Solver~1 uses a uniform (staggered) grid in radius and colatitude, and a second-order finite difference scheme to compute derivatives. We use 100 radial and 90 latitudinal gridpoints, corresponding to  resolutions of $2.3~$Mm and $2^\circ$.} The details of the grid and the computation methods was presented originally in \cite{bekki2022a}. During validation, we found a number of bugs in the code, related to the treatment of the viscous and thermal diffusivities. We use a revised version of the code, where these bugs are corrected for.
In addition, we upgraded the Fortran eigenvalue solver from the \texttt{LAPACK} package to \texttt{ARPACK} \citep[a sparse solver, see][]{lehoucq1998} to speed up the computations.

At fixed $m=1$, we compute the 100 eigenfrequencies nearest to the (real) target frequency of $-90$~nHz 
in the Carrington frame. 
We identify the mode with the largest growth rate (largest imaginary part of the eigenvalue) as the mode of interest. This mode has north-south symmetric radial vorticity, as observed on the Sun. 
Solver~1 gives a mode frequency of $\Re{\omega_{\rm ref}}/2\pi = -95.6$~nHz in the Carrington frame (see solid green line in Fig.~\ref{fig:time_variation}). This is also the only self-excited mode with symmetric radial vorticity in the computed eigenspectrum.

\subsection{Solver~2 from~\citet{mukhopadhyay2025}}
Dedalus \citep{burns2020} is an open-source, parallel code that solves partial differential equations in spectral space. It uses Chebyshev polynomials and spherical harmonics as radial and horizontal basis functions respectively to compute the eigenvalues of a system. Previously, \citet{mukhopadhyay2025} used Dedalus to develop an eigenvalue solver for solar inertial  modes (hereafter Solver~2).
To validate Solver~1, we  used Solver~2 to compute the eigenmodes within the same computational domain. Solver~2 used 84 and 32 gridpoints in radius and colatitude respectively. We specify the same target frequency and compute the same number of eigenmodes as for Solver~1. {The eigenmode with the largest growth rate has a real frequency of $-96.1$~nHz (see dashed line in Fig.~\ref{fig:time_variation}) and a similar eigenfunction to that from Solver~1.}

\subsection{Importance of the position of the lower boundary}
\label{sec:RZ}
As part of our numerical investigations, we found that the inclusion of the upper radiative zone $r=0.65R_\odot~{\rm to }~0.70R_\odot$ in the computational domain plays an important role in unambiguously  identifying the HL1 mode. When considering only the convection zone, as in \cite{mukhopadhyay2025} or \cite{bekki2022a}, the system produces two pairs of self-excited eigenmodes within $\sim 10$~nHz; each pair consisting of two modes with opposite latitudinal symmetries in their eigenfunction. This is at odds with the observation of  a single mode dominating the power spectrum. When the upper layers of the radiative zone are included in the model, however, we find only one pair of self-excited eigenmodes, with the fastest growing mode (eigenfrequency with the largest imaginary part) having the same north-south symmetry as that dominant in the observations. The surface profile of this mode is very similar to that extracted from surface observations. 

\newpage
\section{Additional figures}

\begin{figure}[h]
    \centering
    \includegraphics[width=\linewidth]{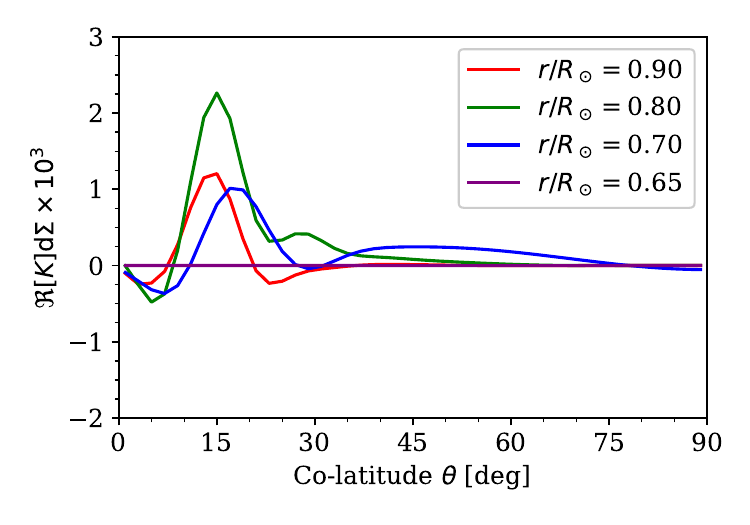}
    \caption{Cuts at different radii though the kernel presented in Fig.~\ref{fig:Kernel}.}
    \label{fig:Kernel_radial_cut}
\end{figure}

\begin{figure}[h]
    \centering
    \includegraphics[width=\linewidth]{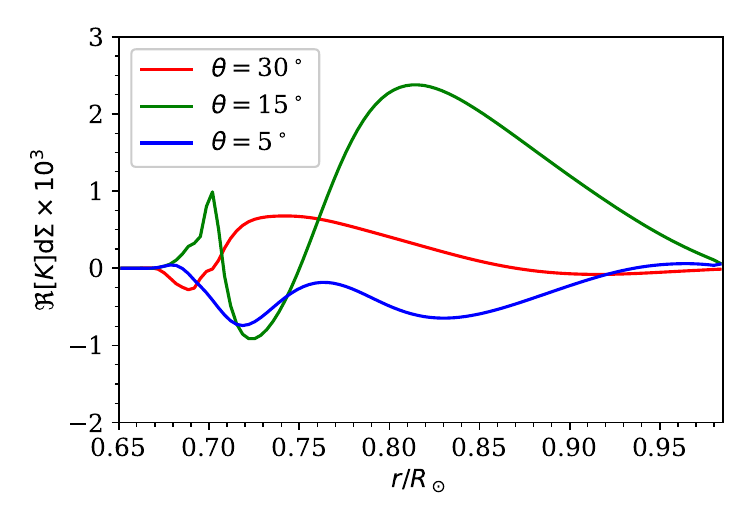}
    \caption{Cuts at different latitudes of the kernel presented in Fig.~\ref{fig:Kernel}.}
    \label{fig:Kernel_latitudinal_cut}
\end{figure}

\end{document}